\documentclass{article}                     
\usepackage[a4paper]{geometry}

\usepackage[ruled,vlined]{algorithm2e}
\usepackage[utf8]{inputenc}
\usepackage{amssymb}
\usepackage{amsmath}
\usepackage{dsfont}
\usepackage[dvipsnames]{color,xcolor}
\usepackage{graphicx}
\usepackage{float, wrapfig}
\usepackage{empheq}
\usepackage{appendix}
\usepackage{enumerate}
\usepackage{mathtools}  
\usepackage{multirow}
\usepackage{tikz}
\usepackage{natbib}
\usepackage{setspace}
\usepackage[affil-it]{authblk} 
\usepackage{hyperref}
\usepackage{ulem}
\usetikzlibrary{shapes,backgrounds,arrows,automata,decorations,shadows,positioning, mindmap}
\newcommand*\xbar[1]{%
   \hbox{%
     \vbox{%
       \hrule height 0.5pt 
       \kern0.5ex
       \hbox{%
         \kern-0.1em
         \ensuremath{#1}%
         \kern-0.1em
       }%
     }%
   }%
} 
\newcommand{\edgeunit}{1.5}
\newcommand{\nodesize}{1em}
\newcommand{\length}{1}
\tikzstyle{covariate}=[draw, rectangle, minimum width=\nodesize, minimum height=\nodesize, inner sep=0, color=black]
\tikzstyle{covmiss}=[draw, minimum width=\nodesize, minimum height=\nodesize, inner sep=0, color=gray, text=gray]
\tikzstyle{observed}=[draw, circle, minimum width=\nodesize, inner sep=0, color=black]
\tikzstyle{basic}=[draw, circle, minimum width=0.4*\nodesize, inner sep=0, color=black, fill=black]
\tikzstyle{clique}=[draw, rectangle, minimum width=2*\nodesize, minimum height=2*\nodesize, inner sep=0, color=black]
\tikzstyle{edge}=[-, line width=1pt, color=black]
\tikzstyle{edgemiss}=[-, line width=1pt, dashed, color=gray]
\tikzstyle{variable}=[scale=0.9,rectangle,draw=white,transform shape,fill=white,font=\Large]

\newcommand{\tr}{\text{tr}}

\newcommand{\betat}{{\widetilde{\beta}}}
\newcommand{\Bt}{{\widetilde{B}}}

\newcommand{\betabf}{{\boldsymbol{\beta}}}
\newcommand{\thetabf}{{\boldsymbol{\theta}}}
\newcommand{\sigmabf}{{\boldsymbol{\sigma}}}
\newcommand{\Omegabf}{{\boldsymbol{\Omega}}}
\newcommand{\Sigmabf}{{\boldsymbol{\Sigma}}}
\newcommand{\Gammabf}{{\boldsymbol{\Gamma}}}
\newcommand{\zerobf}{{\boldsymbol{0}}}
\newcommand{\Xbf}{{\boldsymbol{X}}}
\newcommand{\xbf}{{\boldsymbol{x}}}
\newcommand{\Ybf}{{\boldsymbol{Y}}}

\newcommand{\Wbf}{{\boldsymbol{W}}}
\newcommand{\Ubf}{{\boldsymbol{U}}}
\newcommand{\Mbf}{{\boldsymbol{M}}}
\newcommand{\Qbf}{{\boldsymbol{Q}}}
\newcommand{\Rbf}{{\boldsymbol{R}}}
\newcommand{\Sbf}{{\boldsymbol{S}}}
\newcommand{\mbf}{{\boldsymbol{m}}}

\newcommand\Hcal{{\mathcal{H}}}
\newcommand\Jcal{{\mathcal{J}}}
\newcommand\Ncal{{\mathcal{N}}}
\newcommand\Pcal{{\mathcal{P}}}
\newcommand\Tcal{{\mathcal{T}} }
\newcommand{\Cor}{{\mathds{C}\text{or}}}
\newcommand{\Esp}{{\mathds{E}}}
\newcommand{\Var}{{\mathds{V}}}

\newcommand{\corHTemp}{{\rho(H, \text{temp})}}

\newcommand{\betabft}{{\widetilde{\betabf}}}
\newcommand{\Mbft}{{\widetilde{\Mbf}}}
\newcommand{\Sbft}{{\widetilde{\Sbf}}}

\newtheorem{theorem}{Theorem}
\newtheorem{lemma}{Lemma}

\definecolor{azure}{rgb}{0.0, 0.5, 1.0}
\definecolor{blue2}{rgb}{0.16, 0.32, 0.75}

\title{\textbf{Accounting for missing actors in interaction network inference from abundance data}}

\author{Raphaëlle Momal$^1$%
  \thanks{Electronic address: \texttt{raphaelle.momal@agroparistech.fr}; Corresponding author}, \hspace{0.3cm} Stéphane Robin$^2$, \hspace{0.3cm} Christophe Ambroise$^3$}
\affil{1: UMR MIA-Paris, AgroParisTech, INRAE, Université Paris-Saclay, Paris, France.\\
2: CESCO, Muséum National d'Histoire Naturelle, CNRS, Sorbonne Université, Paris, France.\\
3: Université Paris-Saclay, CNRS, Univ. Évry, Laboratoire de Mathématiques et Modélisation d’Évry 91037, \'Evry, France.}

\date{\today}

\begin{document}

\maketitle
\begin{abstract}
Network inference aims at unraveling the dependency structure relating jointly observed variables. Graphical models provide a general framework to distinguish between marginal and conditional dependency.
Unobserved variables ({\sl missing actors}) may induce apparent conditional dependencies.
In the context of count data, we introduce a mixture of Poisson log-normal distributions with tree-shaped graphical models, to recover the dependency structure, including missing actors. 
We design a variational EM algorithm and assess its performance on synthetic data. We demonstrate the ability of our approach to recover environmental drivers on two ecological datasets.
The corresponding R package is available from \url{github.com/Rmomal/nestor}.\\

\end{abstract}

 \newpage
\section{Introduction} \label{sec:Intro}


\paragraph{Network inference.}
Network inference (or structure inference) has become a topical problem in various fields such as biology, ecology, neuro-sciences, social sciences, to name a few. The aim is to unravel the dependency structure that relates a series of variables that can be jointly observed. Graphical models \citep[see e.g.][]{Lau96} provide a natural framework to achieve this task as it allows to encode the dependency structure into a graph, the nodes of which are the variables. Two variables are connected if and only if they are dependant, conditionally on all others. \\
Most methodologies build on the assumption that the network is sparse, meaning that only a small fraction of variable pairs are conditionally dependent. The case  of Gaussian graphical models (GGM) is especially appealing as the network corresponds to the support of the precision matrix of the joint Gaussian distribution. The use of a sparsity-inducing penalisation gives raise to the celebrated graphical lasso \citep{FHT08}. In a more general context, \cite{ChowLiu} consider a spanning tree structure to impose sparsity to the network, but this drastic form can be alleviated using mixtures of trees \citep{MeilaJaak,kirshner}. \\
One important aspect of network inference is to distinguish between variables that are marginally dependent (possibly because of their respective dependency with some common other) from variables that are {\sl directly related}, that is conditionally dependant. This distinction requires to account for as many confounding effects as possible, which includes all the other variables but also available covariates. It also requires to consider the existence of some {\sl missing actors} (or missing nodes), that may induce an apparent direct dependency.

\paragraph{Abundance data.}
{Count data is found in a multitude of fields (sociology, biology, economy, ecology ...). It results from the counting of events in a given setting such as crime statistics in a state or the number of produced transcripts of a gene in an experiment. The statistical processing of count data cannot always rely on  classical methods developed for continuous Gaussian data and appeals for specific methods. It often exhibits specificities such as zero-inflation and a large dispersion. The present work is motivated by the analysis of so-called abundance data, a count data avatar,  arising from ecological studies where the number of individuals (the abundance) of a series of living species (plants or animals) is observed in a series of sites.} 
In this context, network inference aims at understanding which pairs of species are in direct interaction. The covariates are typically environmental descriptors (altitude, temperature, distance to the see, etc.) of each collection site, while the variables are the respective abundances of each species from the community under study. \\
No nice and generic framework as the GGM exists for count data. A few alternatives rely on copulas  \citep{inouye} or models the node-wise conditional distributions as arising from exponential families. But most joint species distribution models resort to a latent Gaussian layer, which encodes the dependency structure between the species \citep{WBO15,PHW18,PWT19}. The Poisson log-normal model \citep[PLN:][]{AiH89} enters this category: it assumes that a multivariate Gaussian random variable is associated to each species in each site and that the observed abundances are conditionally independent Poisson variables. The PLN model has already been applied to abundance data, both for dimension reduction \citep{CMR18} and network inference \citep{CMR19,MRA20}.

\paragraph{Missing actors.}
In many situations, it is likely that not all actors involved in the system have been observed. The term 'actors' refers to either species that were not observed but nonetheless influence the abundance of others, or environmental conditions that were not accounted for. 

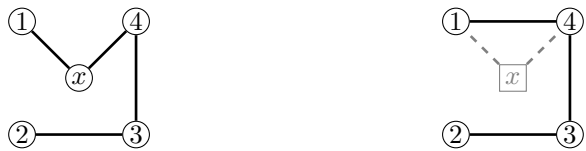
\begin{figure}[H]
 \begin{center}
\begin{tabular}{ccc}
    \begin{tikzpicture}
     \node[observed] (1) at (-0.5*\edgeunit,  .5*\edgeunit) {$1$};
     \node[observed] (2) at (-0.5*\edgeunit, -.5*\edgeunit) {$2$};
     \node[observed] (3) at ( 0.5*\edgeunit, -.5*\edgeunit) {$3$};
     \node[observed] (4) at ( 0.5*\edgeunit,  .5*\edgeunit) {$4$};
     \node[observed] (x) at (0.0*\edgeunit,  .0*\edgeunit) {$x$};
     \draw[edge] (2) to (3); \draw[edge] (3) to (4); 
     \draw[edge] (x) to (1); \draw[edge] (x) to (4);
     \end{tikzpicture}
    &\hspace{3cm} &
    \begin{tikzpicture}
     \node[observed] (1) at (-0.5*\edgeunit,  .5*\edgeunit) {$1$};
     \node[observed] (2) at (-0.5*\edgeunit, -.5*\edgeunit) {$2$};
     \node[observed] (3) at ( 0.5*\edgeunit, -.5*\edgeunit) {$3$};
      \node[observed] (4) at ( 0.5*\edgeunit,  .5*\edgeunit) {$4$};
     \node[covmiss] (x) at (0.0*\edgeunit,  .0*\edgeunit) {$x$};
       \draw[edge] (2) to (3); \draw[edge] (3) to (4); 
    \draw[edgemiss] (x) to (1); \draw[edgemiss] (x) to (4);
     \draw[edge] (1) to (4); 
     \end{tikzpicture}
    \end{tabular}
 \caption{Example of the marginalization when covariate $x$ is unobserved. \textit{Left}: complete graphical model (including $x$). \textit{Right}: marginal graphical model of the observed variables (excluding $x$).}
  \label{fig:MGmissing}
    \end{center}
\end{figure}

In the perspective of unravelling the conditional independence structure, this can typically lead to the inference of spurious edges, which are links between observed actors that are not in direct interaction. 
In the graphical model framework, not accounting for one variable amounts to consider the marginal distribution of the rest of the system, as described in the left panel of Figure \ref{fig:MGmissing}.
Missing actors may be quantitative or qualitative. In the latter case it defines a latent group structure \citep{ambroise2009inferring}.

Several approaches have been proposed for network inference accounting for quantitative missing actors in the context of GGM. 
Many of them \citep{RankSparse,LLVGGM,GirLatent,EMlvggm} adapted the principle of Robust PCA \citep{candes2011robust} to the concentration matrix, assuming it is a sum of two matrices: one low-rank and one sparse. 
In terms of missing actors in a network, the low-rank part corresponds to missing actors connected to all variables, whereas the sparse part refers to missing actors having a local effect.
Following \cite{RAR19} (also in the context of GGM), we focus on the later aspect, that is looking for missing actors not necessarily linked to all others.
As far as we know, no model has been proposed for the inference of missing actors from abundance data.

\paragraph{Variational inference.}
The model we consider in this paper involves different types of variables, namely an unknown tree-shaped graphical model, a continuous latent layer (to induce dependence between the species) and unobserved actors. The most popular approach for the inference of such models is the EM algorithm \citep{DLR77}, which requires the evaluation of the conditional distribution of all unobserved variables given the data. In the problem we consider, some latent variables are (multivariate) continuous and others are discrete, and their joint conditional distribution turns out to be intractable. In this work we resort to a variational approximation \citep{WaJ08} of this conditional distribution and to a  variational EM  algorithm for its inference \citep[see e.g.][]{BKM17}.

\paragraph{Our contribution.}
In the context of the Poisson log-normal model, we propose  a tree-based approach to recover the structure of latent graphical model including actors. The model we consider involves several layers of unobserved variables with intractable conditional distributions, thus we resort to a variational EM algorithm \citep{BKM17} for its inference. We introduce the model in Section \ref{sec:Model} and describe its variational inference in Section \ref{sec:Inference}. The performance of the algorithm is assessed via simulations in Section \ref{sec:Simul}. The use of the proposed model is illustrated in Section \ref{sec:Appli}, where we demonstrate its ability to recover environmental drivers on two ecological datasets. The inference procedure is implemented in the R package nestor, available at \url{github.com/Rmomal/nestor}.

\section{Model} \label{sec:Model}

\subsection{Poisson log-normal and tree-shaped graphical models}

\subsubsection*{Poisson log-normal model.} 
We start with a reminder on the multivariate Poisson log-normal model, with the example of abundance data. The abundances of $p$ species observed on $n$ sites are gathered in the $n \times p$ matrix $\Ybf$ where $Y_ {ij}$ is the count of species $j$ in site $i$, and the row $i$ of $\Ybf$, denoted $\Ybf_i$, is the abundance vector collected on site $i$. A covariate vector $\xbf_i $ with dimension $d$ is also measured on each site $i$ and all covariates are gathered in the $n \times d$ matrix  $\boldsymbol X$. The PLN model states that a (latent) Gaussian vector $\Ubf_i$ of size $p$ with variance matrix $\Rbf = (\rho_{kl})_{kl}$ is associated to each site:
\begin{equation} \label{eq:PLN-Z}
\{\Ubf_i\}_{1 \leq i \leq n} \text{ iid}, \qquad 
\Ubf_1 \sim \Ncal_p(\zerobf, \Rbf),
\end{equation}
the sites being assumed to be independent. To ensure identifiability, we let the diagonal of $\Rbf$ be made of 1's, so $\Rbf$ is actually a correlation matrix.
All latent vectors $\Ubf_i$ are gathered in the $n \times p$ matrix $\Ubf$. The PLN model further assumes that species abundances in all sites are conditionally independent, and that their respective distribution only depends on the environment and the associated latent variable:
\begin{equation} \label{eq:PLN-Y.Z}
\{Y_{ij}\}_{1 \leq i \leq n, 1 \leq j \leq p} \mid \Ubf \text{ independent}, \quad 
Y_{ij} \mid U_{ij} \sim \Pcal\left(\exp(o_{ij} + \xbf_i^\intercal \thetabf_j + \sigma_j U_{ij})\right),
\end{equation}
where $o_{ij}$ is a known offset term which typically accounts for the sampling effort, and $\sigma_j$ is the latent standard deviation associated with species $j$. The vector $d \times 1$ of regression coefficients $\thetabf_j$ describes the environmental effects on species $j$. An important feature of the PLN model is that the sign of the correlation between the observed counts is the same as this of correlation between the latent variables \citep{AiH89}: $\text{sign}(\Cor(Y_{ij}, Y_{ik})) = \text{sign}(\Cor(U_{ij}, U_{ik}))$. 

\subsubsection*{Tree-shaped graphical models.} 
Network inference relies on the assumption that few species are directly dependent on one another, meaning that the underlying graphical model is sparse. In the framework of the PLN model, the graphical model of interest rules the distribution of the latent vectors $\Ubf_i$ and is  encoded in the precision matrix $\Omegabf:=\Rbf^{-1}$. A way to foster sparsity is to impose $\Omegabf$ to be faithful to a spanning tree $T$, that is: $\Ubf_1 \sim \Ncal_p(\zerobf, \Omegabf_T^{-1})$ where the non-zero terms of $\Omegabf_T$ correspond to the edges of the tree $T$ . However this hypothesis is very restrictive  as it allows only $p-1$ links among $p$ species \citep{ChowLiu}. A more flexible approach consists in assuming that the latent vectors are drawn from a mixture of Gaussian distributions, each faithful to a tree $T$ \citep{MixtTrees,MeilaJaak,kirshner,SRS19}:
\begin{equation} \label{eq:mixt-Z}
\Ubf_1 \sim \sum_{T \in \Tcal_p} p(T) \Ncal_p(\zerobf, \Omegabf_T^{-1}),
\end{equation}
where $\Tcal_p$ is the set of spanning trees with $p$ nodes.
We further assume that the tree distribution $\{p(T)\}_{T \in \Tcal_p}$ can be written as a product over the edges:
\begin{equation} \label{eq:prob-T}
p(T) = B^{-1} \prod_{(j, k) \in T} \beta_{jk}, \qquad
\text{with} \quad B = \sum_{T \in \Tcal_p} \prod_{(j, k) \in T} \beta_{jk}.
\end{equation}
The weights $\beta_{jk}$ are gathered in the $p \times p$ symmetric matrix $\betabf$ with diagonal zero. Observe that these weights are defined up to a multiplicative constant, so that only $p(p-1)/2 - 1$ of them may vary independently. This PLN model with latent tree-shaped dependency structure is similar to that considered by \cite{MRA20}.

\subsection{Introducing the missing actor} \label{sec:missActor}

\subsubsection*{PLN model with missing actors.} 
We now introduce the concept of missing actors, which corresponds to variables that are involved in the graphical model but are not associated to observed variables. To involve such actors in the model, we assume that a complete latent vector $\Ubf_i$ with dimension $p+r$ is associated to site $i$, where $r$ is the number of missing actors. This complete vector can be decomposed as $\Ubf_i^\intercal = [\Ubf_{Oi}^\intercal \; \Ubf_{Hi}^\intercal]$ where $\Ubf_{Oi}$ (with dimension $p$) corresponds to observed species and $\Ubf_{Hi}$ (with dimension $r$) corresponds to the missing actors.
The complete $n \times (p+r)$ latent matrix $\Ubf$ can be decomposed in the same way as $\Ubf = [\Ubf_O \; \Ubf_H]$, $\Ubf_O$ and $\Ubf_H$ having dimension $n \times p$ and $n \times r$, respectively. \\ 
The model we consider states that
\begin{enumerate}[($i$)]
\item the complete latent vectors $\Ubf_i$ are all iid and distributed according to a mixture similar to \eqref{eq:mixt-Z} and \eqref{eq:prob-T} but with Gaussian distributions (and matrices $\Omegabf_T$ and $\betabf$) of dimension $(p+r)$, and trees drawn from $\Tcal_{p+r}$;
\item  the abundances $Y_{ij}$ of the  $p$ observed species are distributed according to \eqref{eq:PLN-Y.Z}, replacing $\Ubf$ with $\Ubf_O$,
\end{enumerate}

\begin{figure}[H]
 \begin{center}
	\begin{tikzpicture}	
      \tikzstyle{every edge}=[-,>=stealth',auto,thin,draw]
		\node (A1) at (0.625*\length, 2*\length) {$T$};
		\node (A2) at (0*\length, 1*\length) {$\Ubf_O$};
		\node (A3) at (1.25*\length, 1*\length) {$\Ubf_H   $};
		\node (A4) at (0*\length, 0*\length) {$\Ybf$};
		\draw (A1) edge (A2);
        \draw (A1) edge  (A3);
        \draw (A2) edge  (A3);
        \draw (A2) edge  (A4);
	\end{tikzpicture} 
 \caption{Graphical model linking the count data $\Ybf$, the latent layer of Gaussian parameters $\Ubf=(\Ubf_O,\Ubf_H)$, and the latent tree $T$.}
  \label{fig:MGmodel}
    \end{center}
\end{figure}
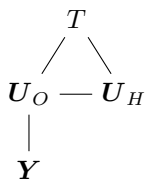

In the sequel, we shall refer to the elements of $\Ubf_O$ and $\Ubf_H$ respectively as 'observed' and 'hidden' (or 'missing') latent variables, whereas obviously none of them are actually observed. Figure \ref{fig:MGmodel} displays the graphical model of the quadruplet $(T, \Ubf_O, \Ubf_H, \Ybf)$. The observed data $\Ybf$ still arise from an PLN model, but the graphical model of the observed latent $\Ubf_O$ may not be sparse due to the marginalization over the hidden latent $\Ubf_H$. Our main goal is to infer the dependency structure of the complete latent vectors, that is to estimate the elements of the matrices $\Omegabf_T$ and the edges weights $\betabf$. The latent dependency structure is similar to this considered by \cite{RAR19}, but the inference strategy much differs, because of the additional hidden layer.

\subsubsection*{Identifiability restriction.} 
The proposed model only makes sense because the graphical model of the complete latent vectors $\Ubf_i^\intercal = [\Ubf_{Oi}^\intercal \; \Ubf_{Hi}^\intercal]$ is supposed to be sparse. Missing actors could obviously not be identified from a regular PLN model, without restriction on the precision matrix $\Omegabf$, as only the marginal precision matrix of the $\Ubf_{Oi}$ could be recovered. Still, to ensure identifiability we impose the same restriction as \cite{RAR19} that  missing latent variables are not connected with each other (the block corresponding to $\Ubf_H \times \Ubf_H$ is diagonal in each $\Omegabf_{T}$).
\section{Inference} \label{sec:Inference}

As said in the introduction, we  resort to a variational EM algorithm to perform the inference due to the complex latent structure.

\subsection{Variational inference}

The log-likelihood of the so-called {\sl complete} data, that is $(\Ybf, \Ubf, T)$, writes
\begin{align*}
\log p_{\thetabf, \betabf, \Omega}(\Ybf, \Ubf, T) 
& = \log p_\betabf(T) + \log p_{\Omegabf}(\Ubf \mid T) + \log p_\thetabf(\Ybf \mid \Ubf)
\end{align*}
where $\Omegabf$ stands for the set of all tree-specific precision matrices: $\Omegabf = \{\Omegabf_T, T \in \Tcal_{p+r}\}$.
The conditional distributions of the latent variables $\Ubf$ and of the tree $T$ given the data $\Ybf$ are both intractable. Variational inference then aims at maximizing a lower bound of the log-likelihood of the observed data, which writes in our context as
\begin{align} \label{eq:lower-bound}
\Jcal(\thetabf, \betabf, \Omega; q)
& = \log p_{\thetabf, \betabf, \Omega}(\Ybf) 
- KL\left(q(\Ubf, T) \| p_{\thetabf, \betabf, \Omega}(\Ubf, T \mid \Ybf) \right) \\
& = \Esp_q \log p_{\thetabf, \betabf, \Omega}(\Ybf,\Ubf, T) + \Hcal(q(\Ubf, T)), \nonumber
\end{align}
where $q(\Ubf,T)$ stands for the approximate joint conditional distribution of the latent layer and of the tree: $q(\Ubf, T) \simeq p(\Ubf, T \mid \Ybf)$. 

\subsubsection*{Approximate distribution.}
The efficiency of variational inference mostly depends on the choice of $q(\Ubf, T)$, which is a balance between computational ease and adequation to the target distribution $p(\Ubf, T \mid \Ybf)$. We adopt here a classical product form for the approximate distribution: we impose to the latent variables $\Ubf$ and to the tree $T$ to be independent according to $q$ (whereas actually they are not conditional on the data), with respective marginals $h$ and $g$:
$$
q(\Ubf, T) =  h(\Ubf)g(T).
$$
Because the sites are independent, and without further assumption, the distribution $h$ is a product over all sites. Following \cite{CMR18} we approximate the conditional distribution of each latent vector $\Ubf_i$ with a Gaussian distribution, that is:
$$
h(\Ubf) = \prod_i \Ncal_{p+r}(\Ubf_i; \mbf_i, \Sbf_i)
$$
with all $\Sbf_i$ diagonal. We gather all the mean vectors $\mbf_i$ in the $n \times (p+r)$ matrix $\Mbf$ and pile up the diagonals of all the variance matrices $\Sbf_i$ in the $n \times (p+r)$ matrix denoted $\Sbf$.

\subsubsection*{Variational EM.}
The variational EM algorithm then consists in maximizing the lower bound $\Jcal$ defined in \eqref{eq:lower-bound} with respect to the parameters (M step), and to the approximate distributions (VE step), alternatively. 
\begin{description}
\item[M step:] At iteration $t+1$, given the current approximate distribution $q^t(\Ubf, T) = g^t(T) h^t(\Ubf)$, the M step consists in the update of the model parameters, solving 
\begin{align} \label{eq:Mstep}
\thetabf^{t+1} & = \arg\max_\thetabf \; \Esp_{h^t} \left[ \log p_\thetabf(\Ybf \mid \Ubf) \right], 
& \Omegabf^{t+1} & = \arg\max_\Omegabf \; \Esp_{q^t} \left[ \log p_{\Omegabf}(\Ubf \mid T) \right], \nonumber \\
\betabf^{t+1} & = \arg\max_\betabf \; \Esp_{g^t} \left[ \log p_\betabf(T) \right].
\end{align}
Observe that the matrix of edge weights $\betabf$ is considered here as a parameter to be estimated, as opposed to \cite{RAR19}, where is was kept fixed and supposed to be given.
\item[VE step:] Maximising $\Jcal$ with respect to (wrt) $q$ is equivalent to minimizing the K\"ullback-Leibler divergence between $q(\Ubf, T)$ and $p_{\thetabf, \betabf, \Omega}(\Ubf, T \mid \Ybf)$ that appears in \eqref{eq:lower-bound}. Because we adopted a product form for $q$, the solution of the VE step for both $g$ and $h$ is known to be a mean-field approximation \citep{WaJ08}. More specifically, maximising $\Jcal$ gives
\begin{align} \label{eq:VEstep-g}
g^{t+1}(T) 
& \propto \exp \left\{ \Esp_{h^t} \left[ \log p_{\thetabf^{t+1}, \betabf^{t+1}, \Omega^{t+1}}(\Ybf, \Ubf, T) \right] \right\} \nonumber \\
& \propto \exp \left\{ \log p_{\betabf^{t+1}}(T) + \Esp_{h^t} \left[ \log p_{\Omegabf^{t+1}}(\Ubf \mid T) \right] \right\},
\end{align}
and
\begin{align} \label{eq:VEstep-hH}
h^{t+1}(\Ubf) 
& \propto \exp \left\{ \Esp_{g^{t+1}} \left[ \log p_{\thetabf^{t+1}, \betabf^{t+1}, \Omega^{t+1}}(\Ybf, \Ubf, T) \right] \right\} \nonumber \\
& \propto \exp \left\{ \Esp_{g^{t+1}} \left[ \log p_{\Omegabf^{t+1}}(\Ubf \mid T) \right] + \log p_{\thetabf^{t+1}}(\Ybf \mid \Ubf) \right\}. 
\end{align}
\end{description}

Observing that $\log p_\betabf(T) + \log p_{\Omegabf}(\Ubf \mid T)$ can be written as a sum over all the edges present in $T$, we see that $g^{t+1}(T)$ has a product form. So, without any further assumption, we may parametrize $g(T)$ in the same way as $p_\betabf(T)$:
\begin{equation} \label{eq:g}
g(T) = \prod_{jk \in T} \betat_{jk} / \Bt
\qquad \text{where} \quad
\Bt = \sum_{T \in \Tcal_{p+r}} \prod_{jk \in T} \betat_{jk}.
\end{equation}
We gather the $\betat_{jk}$'s in the $(p+r) \times (p+r)$ matrix $\betabft$. The parameters $\betabft$, $\Mbf$ and $\Sbf$ are called the variational parameters, in the sense that it is equivalent to optimize $\Jcal$ wrt $(g, h)$ or wrt $(\betabft, \Mbf, \Sbf)$.

\subsection{Proposed algorithm}
\label{algo}

The model we consider is an extension of the PLN model, for which an efficient inference algorithm have been implemented in the \url{PLNmodels}, an R package available on CRAN \citep{CMR18,CMR19}. 

\subsubsection*{Prior estimates of $\thetabf$, $\Mbf_O$ and $\Sbf_O$.}
To alleviate the computational burden of the inference, we take advantage of this available tool to get an estimate of the regression coefficient matrix $\widehat{\thetabf}$ and an approximation of the parameters  of the observed latent variable conditional distribution $h_O(\Ubf_O) \simeq p(\Ubf_O \mid \Ybf)$. These latter parameters are $\Mbf_O$ and $\Sbf_O$ (first $p$  columns of $\Mbf$ and $\Sbf$ respectively) and we denote $\Mbft_O$ and $\Sbft_O$ their approximation. The quantities $\widehat{\thetabf}$, $\Mbft_O$ and $\Sbft_O$ are kept fixed in the rest of the algorithm, so the VEM algorithm only deals with the remaining unknown quantities: the model parameters $\betabf$, $\Omegabf$, and the variational parameters $\betabft$, $\Mbf_H$, $\Sbf_H$. 
As a consequence, the final estimates we get yield a lower value of the objective function $\Jcal$ as compared to  an optimisation wrt to all model and variational parameters.

\subsubsection*{M step.} This steps deals with the update of the  model parameters $\betabf$ and $\Omegabf_T$. Some of the calculations are tedious and postponed to Appendix \ref{app:comput}.

\paragraph{Edges weights $\betabf$:} 
As shown in Equation \eqref{eq:Mstep}, the maximization of $\mathcal{J}$ requires the computation of the derivative of $\Esp_{g^t} [\log p_{\betabf}(T)]$ wrt $\betabf$, which includes the derivative of the normalizing constant $B$. The latter can be computed via an extension of the Matrix Tree theorem \citep[see][Lemma \ref{lem:Meila} reminded in Appendix \ref{app:tools}]{MeilaJaak}. Setting the derivative of the expectation to 0 yields the following update (same as in \citet{MRA20} and detailed in appendix \ref{up:beta}):
$$
\beta^{t+1}_{kl} 
= \frac{P^t_{kl}}{ M(\betabf^t)_{kl}},
$$
where $M(\betabf)$ is defined in Lemma \ref{lem:Meila} and $P^t_{kl}$ is the probability that the edge $(k, l)$ belongs to the tree $T$ according to $g^t$:
$$
P^t_{kl} = \mathds{P}_{g^t}\{kl \in T\} 
= \sum_{\substack{T  \in \mathcal{T}: \\ T \ni kl}} g^t(T) 
= \frac{1}{\Bt^t} \sum_{\substack{T  \in \mathcal{T}: \\ T \ni kl}} \prod_{uv \in T} \betat^t_{uv}.
$$
$P^t_{kl}$ is computed using a result from \citet{kirshner} (reminded as Lemma \ref{lem:Kirshner} in appendix A). We now define the binary variable $I_{Tkl}$ which indicates the presence of the edge $kl$ in tree $T$, so $P_{kl}^t = \Esp_{g^t} [I_{Tkl}]$ and $I_T = [I_{Tkl}]_{1 \leq k, l \leq (p+r)}$ is the adjacency matrix of tree $T$.

\paragraph{Precision matrices $\Omegabf_T$:}
For a given dependency structure in the Gaussian Graphical model framework, \cite{Lau96} gives maximum likelihood estimates for the precision matrix. 
These estimators are given as functions of sufficient statistics of the multivariate Gaussian distribution. Indeed in the exponential family framework, the M step of any EM algorithm requires the computation of the expectation of a sufficient statistic, under the current fit of the variational laws (see \citet{mclachlan}). Here as $\Ubf \mid T$ is centered, a sufficient statistic is $\Ubf^\intercal \Ubf$. We now let $SSD$ denote the matrix defined as 
$$
SSD^t = \Esp_{h^t} (\Ubf^\intercal \Ubf) = (\Mbf^t)^\intercal \Mbf^t + \Sbf^t_+
$$
where $\Sbf^t_+ = \sum_i \Sbf^t_i$. Applying Lauritzen's formulas, we get:
\begin{align} \label{omegaT}
\omega^{t+1}_{Tkl} & = \left\{
\begin{array}{ll}
\dfrac{ -ssd_{kl}^{\,t}/n}{1-(ssd_{kl}^{\,t}/n)^2} & \text{if } kl \in T \\
0 & \text{otherwise}
\end{array} 
\right., \\
\omega^{t+1}_{Tkk} & = 1 + \sum_l I_{Tkl} \dfrac{(ssd_{kl}^{\,t}/n)^2}{1-(ssd_{kl}^{\,t}/n)^2},
\nonumber
\end{align}
where $ssd^t_{kl}$ stands for the entry $kl$ of the matrix $SSD^t$.
The calculations are postponed to Appendix \ref{up:omega}. Observe that estimates of the off-diagonal entries $\omega^{t+1}_{Tkl}$ do not depend on $T$ provided that the edge $(k, l)$ belongs to $T$.  Thus the estimates of the off-diagonal terms of the precision matrices $\Omegabf_T$ are common to all trees sharing a given edge. This does not result from any assumption on the shape of  $\Omegabf_T$, but from the properties of the maximum likelihood estimate of Gaussian variance matrix. In the sequel we will simply denote off-diagonal terms by $\omega_{kl}$ (as opposed to $\omega_{Tkk}$ which still depends on $T$).\\

Other quantities are needed for later computations. Lauritzen  gives the maximum likelihood estimator of every entry of the correlation matrix $\Rbf_T$ corresponding to an edge $kl$ being part of $T$, which is
$ \Rbf_{Tkl}^{t+1} = ssd_{kl}^{\,t}/n. $
Hereafter for any matrix $A$, $A_{[kl]}$ refers to the bloc $kl$ of $A$: $A_{[kl]}=(a_{ij})_{\{i,j\}\in\{k,l\}}$. The determinant of $\Omegabf^{t+1}_T$ factorizes on the edges of $T$ and writes as a function of blocs of the correlation matrix  as follows:  

\begin{align}
    |\Omegabf^{t+1}_{T}| = \Big(\prod_{kl \in T} |\Rbf_{T[kl]}^{t+1}|\Big)^{-1}
\quad 
\text{and for any $kl \in T$, } 
|\Rbf_{T[kl]}^{t+1}|= 1-(ssd_{kl}^{\,t}/n)^2. \label{RT}
\end{align}

Finally we define the matrix  $\xbar{\Omegabf}^{\,t+1} = \Esp_{g^t}[\Omegabf^{t+1}_T]$. 
Noticing that, for $k \neq l$, $\Esp_{g^t}[\Omegabf^{t+1}_T]_{kl} = \Esp_{g^t}[\Omegabf^{t+1} \odot I_T]_{kl}$, edges probabilities appear as follows:
$$
\xbar{\omega}_{kl}^{\,t+1} 
= - P_{kl}^t\dfrac{ ssd_{kl}^{\,t}/n}{1-(ssd\,^t_{kl}/n)^2}, 
\qquad 
\xbar{\omega}_{kk}^{\,t+1}= 1+\sum_l P_{kl}^t \dfrac{(ssd_{kl}^{\,t}/n)^2}{1-(ssd_{kl}^{\,t}/n)^2}.
$$

\subsubsection*{VE step.} 
This step deals with the update of the approximate conditional distributions $g$ and $h_H$, namely the update of the corresponding variational parameters $\betabft$, $\Mbf_H$ and $\Sbf_H$.

\paragraph{Approximate conditional tree distribution $g(T)$:}
Computing the expression \eqref{eq:VEstep-g} yields the following, where the constant term 'cst' does not depend on a specific edge:
\begin{align*}
 \log g^{t+1}(T) &= \log p_{\betabf^{t+1}}(T) + \Esp_{h^t} \left[ \log p_{\Omegabf^{t+1}}(\Ubf \mid T) \right]  + \text{cst}  \\
&=  \sum_{kl \in T} \log \beta^{t+1}_{kl} 
- \frac{n}2 \log |\Rbf_{[kl]}^{t+1} | 
-  \omega^{t+1}_{kl} \left[(\Mbf^t)^\intercal \Mbf^t\right]_{kl} + \text{cst}
\end{align*}
  Then remembering the product form of  $g^{t+1}$ given in \eqref{eq:g}, we obtain the expression for each edge variational weight:
  \begin{equation}
      \betat^{t+1}_{kl} = \beta^{t+1}_{kl} \left|\Rbf_{[kl]}^{t+1}\right|^{-n/2} \exp\left(-\omega^{t+1}_{kl} \left[(\Mbf^t)^\intercal \Mbf^t\right]_{kl}\right).
  \end{equation}

%

\paragraph{Approximate Gaussian distribution $h$:} 
According to \eqref{eq:VEstep-hH}, we have that
$$
\log h^{t+1}(\Ubf) 
= \Esp_{g^{t+1}} \log p(\Ybf \mid \Ubf_O) - \frac12 \tr\left(\xbar{\Omegabf}^{t+1}_T (\Ubf^\intercal \Ubf)\right) + \text{cst}.
$$
Using the properties of the conditional Gaussian distribution we have that
$$
h^{t+1}(\Ubf_H \mid \Ubf_O) = \Ncal\left(\Ubf_H; 
-\Ubf_O \xbar{\Omegabf}^{t+1}_{OH} \left(\xbar{\Omegabf}^{t+1}_{H}\right)^{-1}, \left(\xbar{\Omegabf}^{t+1}_{H}\right)^{-1}\right).
$$
Now, to get $h_H^{t+1}(\Ubf_H)$, it suffices to integrate $h^{t+1}(\Ubf_H \mid \Ubf_O)$ wrt $h_O$ (the parameter of which are kept fixed along iterations) to get
$$
\Mbf^{t+1}_H = -\widetilde{\Mbf}_O \xbar{\Omegabf}^{t+1}_{OH} \left(\xbar{\Omegabf}^{t+1}_{H}\right)^{-1}, 
\qquad
\Sbf^{t+1}_H = \left(\xbar{\Omegabf}^{t+1}_{H}\right)^{-1}.
$$


\subsection{Algorithm peculiarities} \label{sec:algoSpec}
\subsubsection*{Initialization.}
\label{init}
As for any EM algorithm, the choice of the starting point is paramount. The initialization we use here takes the primary estimate $\widetilde{M}_O$ as an input.
\begin{description}
\item [Initial clique:]
As a starting point, we look for a clique of species as potential neighbors of the missing actor $h$. There are many different ways to do so, and if any prior knowledge exists on that matter it should be used. Otherwise, such a clique can be found using sparse principal component analysis \citep[sPCA;][]{spca}, where principal components are formed using only a few of the original variables, which is consistent with the assumption that each missing actor is connected only to some actors in the network. 

When applying sPCA to $\widetilde{M}_O$, the set of non-zero loadings of each principal components provides us with an initial clique of neighbors of each missing actor

\item[Parameters initialization:]
The eigenvectors resulting from the sPCA also provide us with a starting value $\Mbf^0_H$, as well as a first estimate of the latent correlation matrix $\Rbf^0$. The parameter $\betabf$ is uniformly initialized.
\end{description}
 
\subsubsection*{Numerical issues.}

Because the Matrix Tree Theorem and Kirshner's formula respectively resort to the calculation of a determinant and a matrix inversion, the proposed algorithm is exposed to numerical instabilities. To circumvent these issues, we rely on both multiple-precision arithmetic and likelihood tempering \citep[via a parameter $\alpha$, similarly to][]{ScR17}. More details are given in Appendix \ref{app:numIssues}.

\section{Simulations} \label{sec:Simul}
\subsection{Count datasets}
For the simulation study, 300 count datasets of $15$ species in total including one missing actor are generated, thus $p=14$ and $r=1$. 
Data is generated as follows.
We generate a scale-free structure $\mathcal{G}$ (which degree distribution is a power law) with $p+1$ nodes using the R package \texttt{huge} \citep{zhao2012huge} available on CRAN. 
The missing species $h$ is chosen as the one with highest degree. We measure the {\sl influence} of the missing actor with its degree, distinguishing three influence classes: \textit{Minor} (degree $\leq 5$), \textit{Medium} ($5<$ degree $\leq 7$) and \textit{Major} (degree $\geq 8$). For each replicate, the latent layer $\Ubf$ and the observed abundances $\Ybf$ are simulated according to the model defined in Section \ref{sec:Model}.

\subsection{Experiment \& Measures}
For each simulated dataset, the VEM algorithm is initialized as described in Section \ref{init}. 
More specifically and because we only look for one missing actor, we consider the cliques corresponding to each of the first two principal components of sPCA, and their respective complements, which provides us with four cliques.
Then four VEM algorithms, as described in Section \ref{algo}, are run starting from each of the four candidate cliques, and the one yielding the highest lower bound $\Jcal$ is kept. 
For all simulations, we set the precision of the convergence criterion to $\varepsilon=10^{-3}$, the tempering parameter to $\alpha=0.1$ and the maximal number of iterations to $100$. 
The  inference quality is assessed regarding the global network inference, the missing actor's position in the network, and its values along the $n$ sites. We refer to this first procedure as the \textit{blind} procedure. Additionally, we define the \textit{oracle} procedure as running the VEM with the set of true neighbors of the missing actor as initial clique.\\

For each procedure, a general measure of the whole network inference quality is first given by comparing the inferred edge probabilities to the original dependency structure. This is done using the Area Under the ROC Curve (AUC) criteria.  Then, to be more specific and target the neighbors of node $h$ specifically, the probabilities of edges involving $h$ are transformed into binary values using the 0.5 threshold. The values are then compared to the original links of $h$ and yield quantities of true/false ($T$/$P$) positives/negatives ($P$/$N$), from which are built the \textit{precision} (also known as the positive predictive value, ${TP}/({TP+FP)}$) and the \textit{recall} (also known as the true positive rate, ${TP}/({TP+FN})$) criteria. Finally, we assess the ability to reconstruct the missing actor across the sites  by computing the absolute correlation between its inferred vector of means ($M_h$) and its original latent Gaussian vector $\Ubf_h$.

\subsection{Results}
\begin{figure}[H]
    \centering
    \includegraphics[width=11cm]{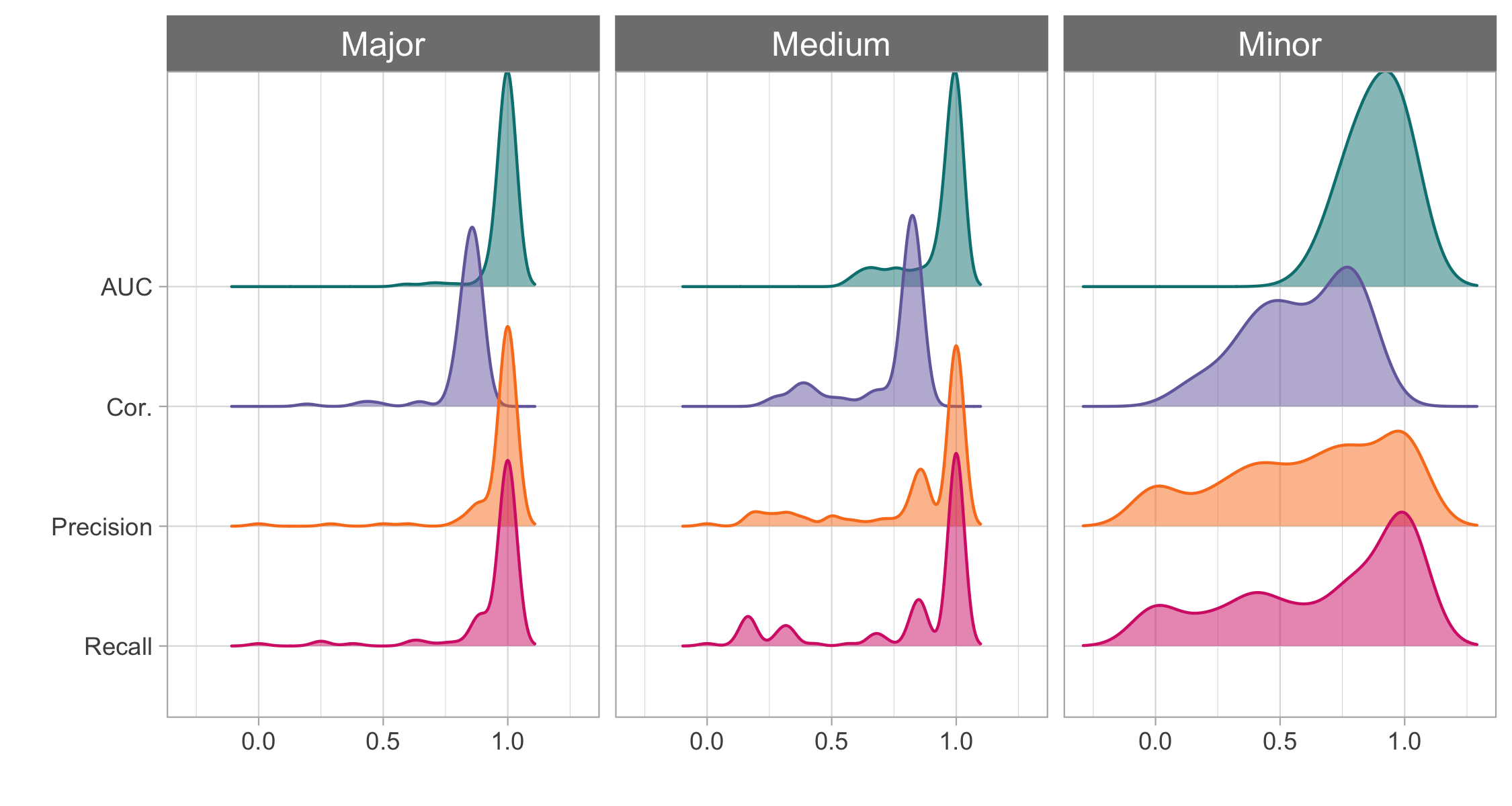}
    \caption{
    The influence of the missing actor is measured with its degree, distinguishing three influence classes: \textit{Minor} (degree $\leq 5$), \textit{Medium} ($5<$ degree $\leq 7$) and \textit{Major} (degree $\geq 8$). 
    The distributions of performance measures are displayed for each class of influence: AUC measures the retrieval of the dependence structure between all variables, observed and missing.  Precision and recall are specific to the missing actor links. 
    }
    \label{fig:densities}
\end{figure}

Simulations performance measures are gathered in Table \ref{tab:perf} and Table \ref{tab:oracle} for blind and oracle procedures respectively. The distributions of the quality measures are displayed in Figure \ref{fig:densities}. \\

Table \ref{tab:perf} shows the network is well inferred, as all AUC means are above 0.85, with almost perfect inference when the influence of the missing actor is major. Its neighbors and values per site are very well retrieved in these cases with mean recall values above 0.9 and mean correlation above 0.8, with a great confidence in the algorithm outputs as mean precision is above 0.95. However, there exists a clear deterioration of all performance as the influence decreases with lower means are greater deviations, down to about 0.6 mean values for all measures when the influence is minor. Moreover, the algorithm takes more and more time to converge as the influence decreases, although it stays at about $3s$ for minor cases which is reasonable. Figure \ref{fig:densities} shows that as the influence decreases, the densities present with several modes and dilute towards 0, illustrating that even if some networks are still well-inferred, there also are more and more cases where the algorithm fails. In particular, the performance decrease of medium cases seems to be only due to a greater number of failed inferences.\\

All these elements point to minor cases being harder problems to solve, unsurprisingly. Yet as oracle results show in Table \ref{tab:oracle}, it is possible to carry out almost-perfect inference in all cases, if the algorithm is initialized with the true clique; the deterioration is still present in all measures, but stays marginal. Thus the harsh decrease in the blind procedures seems to be mainly due to the proposed initialization method failing at correctly finding some of the small cliques of neighbors.\\

\begin{figure}[H]
    \centering    \includegraphics[width=11cm]{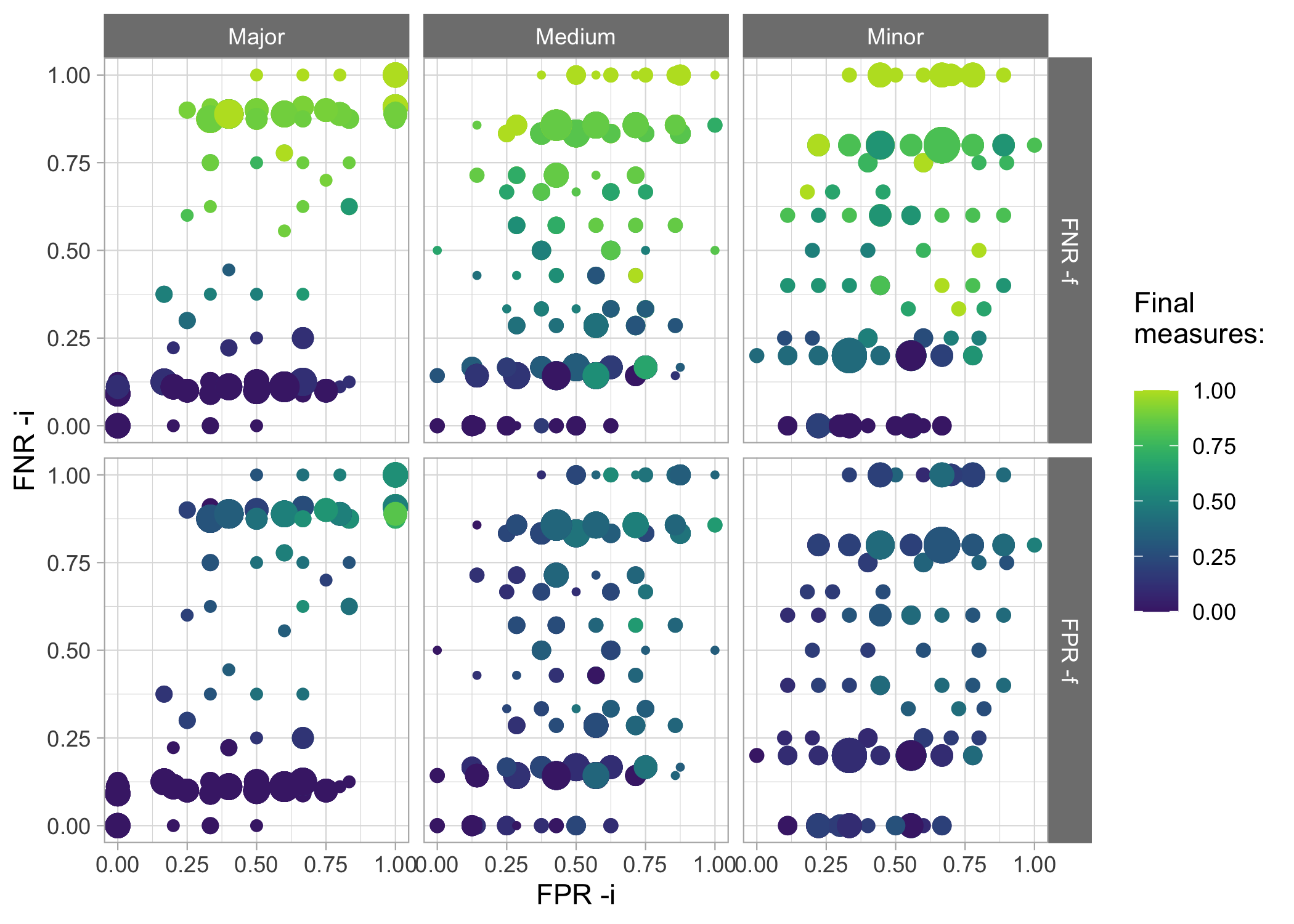}
    \caption{Comparison of initial and final FPR and FNR, for cliques of neighbors of one missing actor obtained with the sparse PCA method. Position of dots are defined according to initial values, their color according to the final FPR and FNR. Sizes are proportional to the density of dots on a given position.}
    \label{fig:perfinit}
\end{figure}

\paragraph{About intialization.}
Figure \ref{fig:perfinit} compares the initialization quality and the corresponding final inferred neighbors, in terms of initial (-i) and final (-f) false negative (FNR, also 1-TPR) and positive rates (FPR). It clearly appears that final measures  mostly increase with false negatives of the initial clique. This means that not including a neighbor in the initialization is much worse for the inference than falsely including a node. The increase of FNR-f is bigger than that of FPR-f, meaning that a wrong initialization leads to a set of inferred neighbors which most part can be trusted, but which will be largely incomplete. This advocates for bigger initialization cliques when no prior information is available.

\begin{table}
\centering
\begin{tabular}{lrrrrrr}
  \hline
  & N & AUC & Precision & Recall & Correlation & Time (s) \\ 
  \hline
 Major & 100 & 0.98 (0.06) & 0.96 (0.14) & 0.94 (0.17) & 0.83 (0.10)& 2.36 (0.91)  \\ 
 Medium & 132 & 0.93 (0.12) & 0.83 (0.26) & 0.81 (0.30) & 0.73 (0.17)& 2.69 (1.15)  \\ 
 Minor &  68 & 0.89 (0.10) & 0.61 (0.34) & 0.66 (0.36) & 0.59 (0.21) & 3.08 (1.14) \\ 
   \hline
\end{tabular}
\caption{\label{tab:perf}Blind procedure using cliques from initialization. The influence of the missing actor is measured with its degree, distinguishing three influence classes: \textit{Minor} (degree $\leq 5$), \textit{Medium} ($5<$ degree $\leq 7$) and \textit{Major} (degree $\geq 8$).  For each class of influence, the following quantities are reported:  number of simulated graphs (N), means and standard deviations of AUC, Precision, Recall, Correlation between missing actor inferred vector of means and original latent vector, and running times in seconds. AUC measures the retrieval of the dependence structure between all variables (observed and missing), whereas precision and recall are specific to the missing actor links.} 

\end{table}

 \begin{table}
\centering
\begin{tabular}{lrrrrrr}
  \hline
  & N & AUC & Precision & Recall & Cor. &t(s) \\ 
  \hline
  Major & 100 & 1 (0.00) & 1 (0.00) & 1 (0.01) & 0.86 (0.02)  & 1.28 (0.21) \\ 
  Medium & 132 & 1 (0.02) & 1 (0.00) & 0.99 (0.04) & 0.83 (0.02)  & 1.38 (0.46)  \\ 
  Minor &  68 & 0.98 (0.04) & 0.99 (0.03) & 0.96 (0.12) & 0.8 (0.04)& 1.56 (0.69) \\ 
   \hline
\end{tabular}
 \caption{\label{tab:oracle} Oracle procedure using true clique as starting point. The influence of the missing actor is measured with its degree, distinguishing three influence classes: \textit{Minor} (degree $\leq 5$), \textit{Medium} ($5<$ degree $\leq 7$) and \textit{Major} (degree $\geq 8$).  For each class of influence, the following quantities are reported:  number of simulated graphs (N), means and standard deviations of AUC, Precision, Recall, Correlation between missing actor inferred vector of means and original latent vector, and running times in seconds. AUC measures the retrieval of the dependence structure between all variables (observed and missing), whereas precision and recall are specific to the missing actor links.}
\end{table}
\section{Applications}  \label{sec:Appli}

\subsection{Cross validation criterion for  model selection}
%

The proposed model obviously raises the problem of choosing the number of missing actors $r$ (which may be zero). Variational-based inference often relies on approximate versions of the BIC or ICL criteria for model selection. Few theoretical guaranties exist about these approximate criteria and, in the present case, we observed that BIC and ICL penalizations did not yield consistent results. Therefore, we resort to $V$-fold cross validation to determine the number of missing actors. 

More specifically, we split the original dataset $\Ybf$ ($\Xbf$ is dropped here for the sake of clarity) into $V$ subsets with almost equal sizes $m_1, \dots m_V$ ($\sum_{v=1}^V m_v = n$), which we denote $\{\Ybf^v\}_{v = 1, \dots V}$. For each subset $v$, we define its complement $\Ybf^{-v}$ on which we fit a model with $r$ missing actors and get a parameter estimate $\Gammabf_r^{-v} = (\thetabf_r^{-v}, \sigmabf_r^{-v}, \betabf^{-v}_r, \Omegabf_r^{-v})$ and measure the fit of $\Gammabf_r^{-v}$ to the test dataset $\Ybf^v$. 

To avoid the integration over the $(p+r)$-dimensional Gaussian latent layer, we measure the fit with the pairwise composite likelihood \citep{lindsay}.
For any given tree $T$ and parameter $\Gammabf$, the bivariate Poisson log-normal pdf $p_{PLN}\left((Y_{ij}, Y_{ik}); \Gammabf, T \right)$ can be easily computed for any sample $i$ and pair of species $(j, k)$ with available tools such as the \texttt{poilog} R package \citep{ViS08} available on CRAN. The cross-validation criterion is defined as
$$
PCL_r(\Ybf) = \frac1V \sum_v \frac1B \sum_{b=1}^B \frac1{m_v} \sum_{i = 1}^{m_v} \sum_{j < k} \log p_{PLN}\left((Y^v_{ij}, Y^v_{ik}); \Gammabf_r^{-v}, T_{r, b}^{-v} \right)
$$
where the tree samples $\{T_{r, b}^{-v}\}_{b=1 \dots B}$ are iid according to $p_{\betabf_r^{-v}}(T)$. 

The sampling procedure for spanning trees is given in Appendix \ref{eq:sampTree}; the complete procedure for the calculation of $PCL_r(\Ybf)$ is described by Algorithm \ref{algo:model-selection}, given in Appendix \ref{sec:modSel}. Note that this criterion measures the fit of the model in terms of abundance prediction, whereas our interest is mostly focused on the inference of the dependency structure. In other words, our goal is identification, that is selecting the smallest model  and not the best model in terms of prediction \citep{arlot2010survey}.

\begin{figure}[H]
    \centering
    \includegraphics[width=10cm]{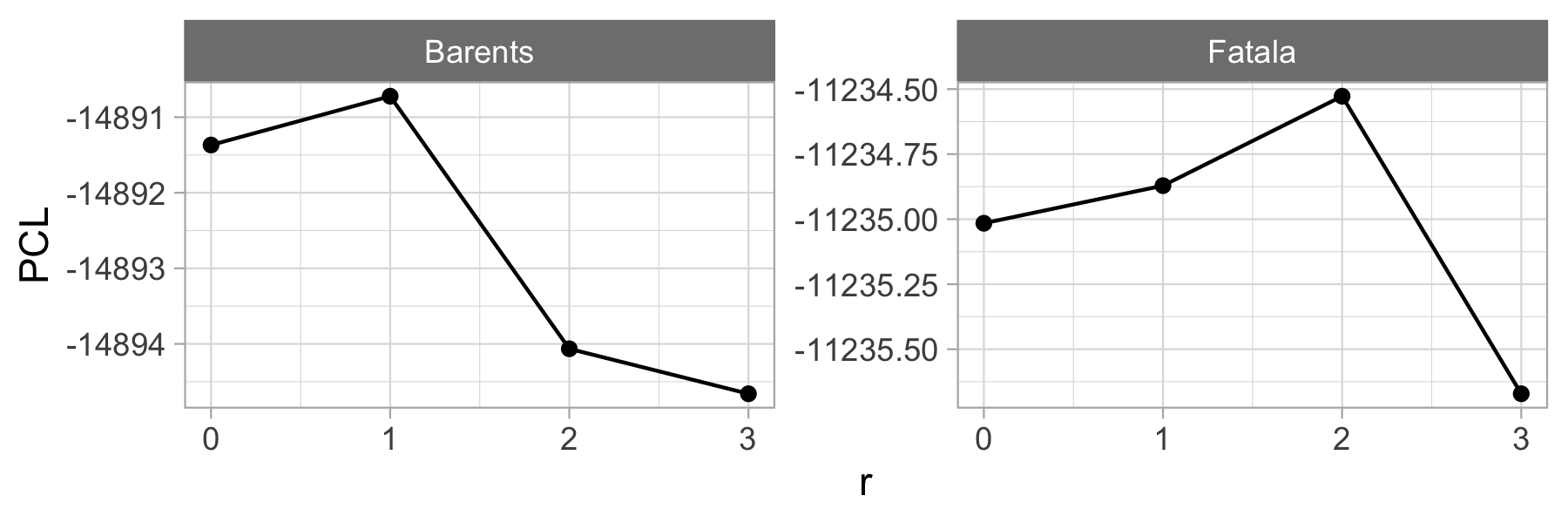}
    \caption{Pairwise composite likelihoods estimates of Barents and Fatala datasets for models including 0 to 3 missing actors.}
    \label{fig:selec}
\end{figure}

We did not include this computationally greedy procedure in the simulation study but applied it to the two ecological datasets that will be described in the next two sections. The results, gathered in Figure \ref{fig:selec}, yield $r=1$ missing actor for the Barents Sea data set, and $r=2$ missing actors for the Fatala River one.

Regarding the initialization, we performed a wider exploration as compared to the simulation study. To enlarge the list of possible cliques, we applied a resampling version of the procedure described in Section \ref{sec:algoSpec}, and applied it to 200 sub-samples, each consisting in 80\% of the whole data set. This yielded 200 lists of $r$ initial cliques, from which duplicates were removed.

\subsection{Barents Sea}

The dataset was first published by \cite{FNA06} and consists of the abundance of 30 fish species measured in 89 sites in the Barents See in April-May 1997. In addition to abundances, the water temperature was measured in each site. The complete dataset is available at \url{www.fbbva.es/microsite/multivariate-statistics/data.html}. 
Fishes distributions are known to be greatly linked with the temperature. 
Hence to illustrate our methodology, 
we present the results of the model fitted without any covariate (that is not accounting for the temperature), but including one missing actor (as suggested by Figure \ref{fig:selec}). To assess the ability of the proposed methodology to retrieve the influence of temperature as a missing actor, we report the empirical correlation between the temperature and the conditional expectation of the missing actor $M_h$, which we denote $\corHTemp$.
 
The resampling initialization procedure yielded in 14 different cliques, for each of which a VEM algorithm was run: the mean running time was $6.63$mins with deviation $0.70$ mins. 

\begin{figure}[H]
    \centering
    \includegraphics[width=10cm]{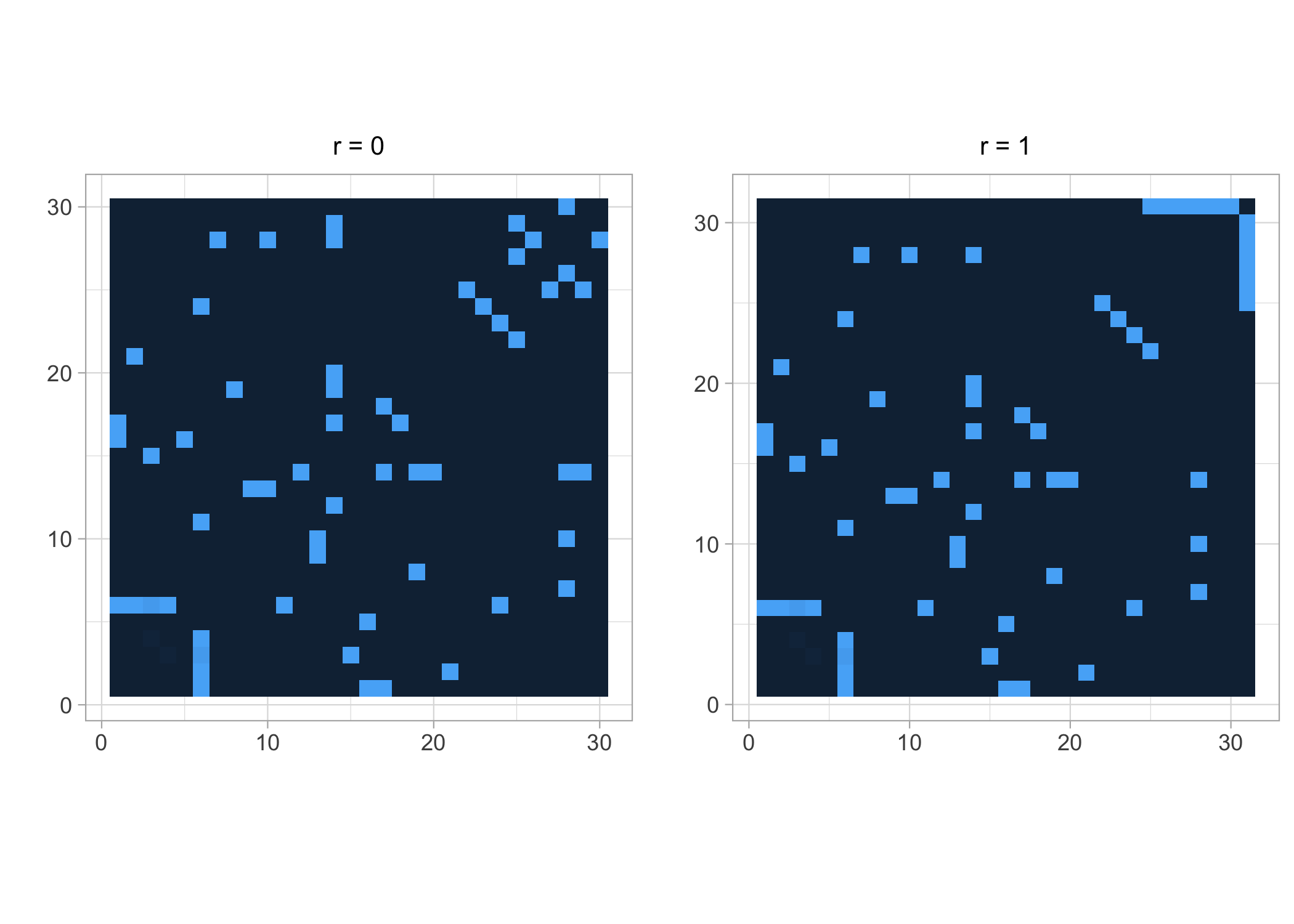} \\\vspace{-2cm}
    \includegraphics[width=10cm]{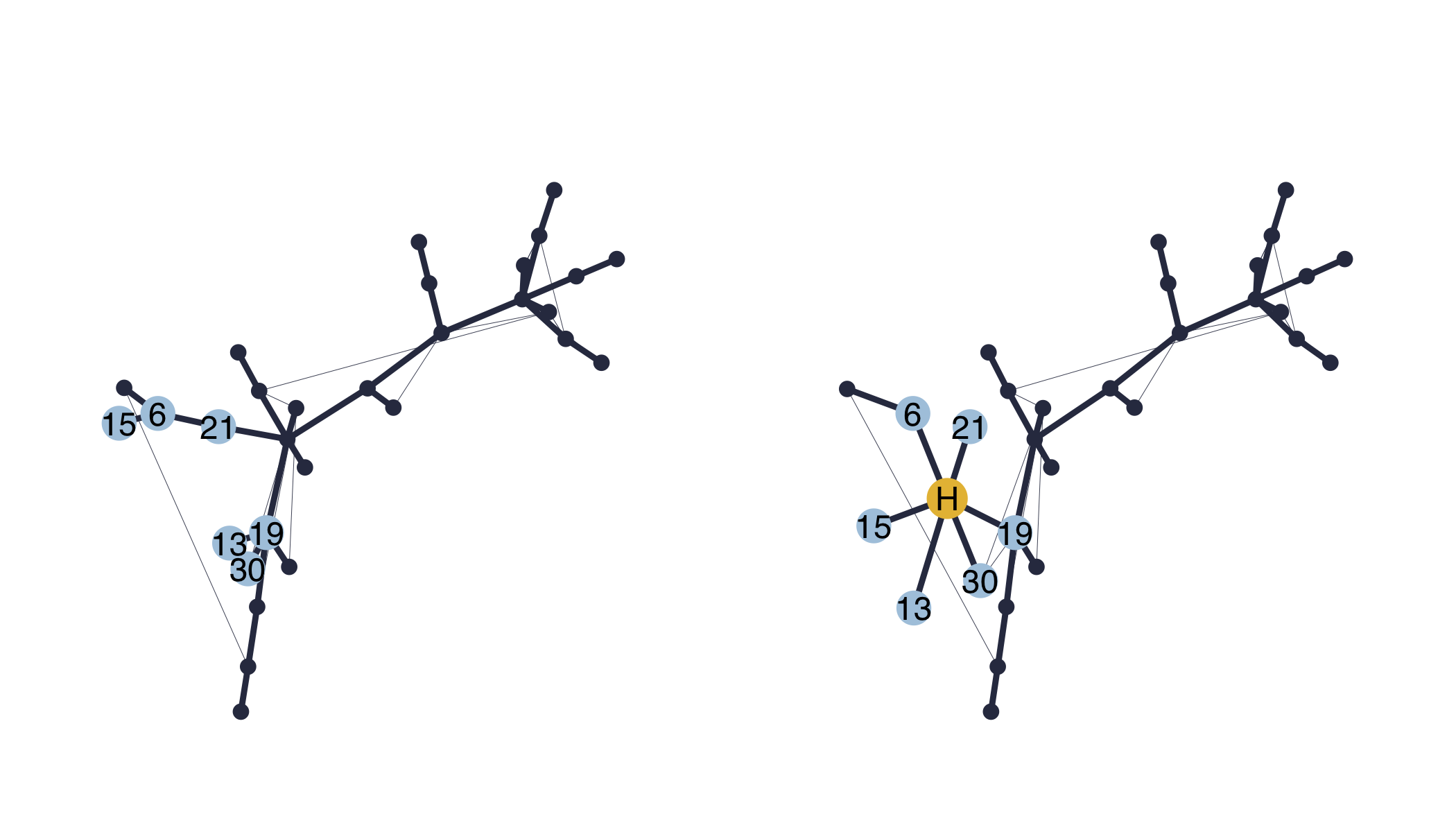}
    \caption{\textit{Top left:} adjacency matrix of the Barents Sea fishes interaction network for $r=0$  missing actor. The inferred neighbors are gathered in the last 6 columns, so that their interactions are observable in the upper-right corner. \textit{Top right:} adjacency matrix for $r=1$  missing actor. The last column gathers the interactions of the inferred missing actor. \textit{Bottom}: Inferred interaction network with $r=0$ (left) and $r=1$ (right). Colored nodes refer to the inferred neighbors (blue) of the missing actor (yellow). The edges width are proportional to their probability.}
    \label{fig:barents_adj}
\end{figure}

The edge probabilities involving node $h$ as an endpoint were either very close to 0 or very close to 1, yielding a total of 6 highly probable neighbors of $h$. Figure \ref{fig:barents_adj} shows that many direct interactions are inferred between the corresponding 6 species in absence of a missing actor, which vanish when it is introduced. It also shows that accounting for this actor has only a local effect and that the direct interactions among the other species are preserved, which is consistent with our notion of a missing actor.

\begin{figure}
    \centering
    \includegraphics[width=5cm]{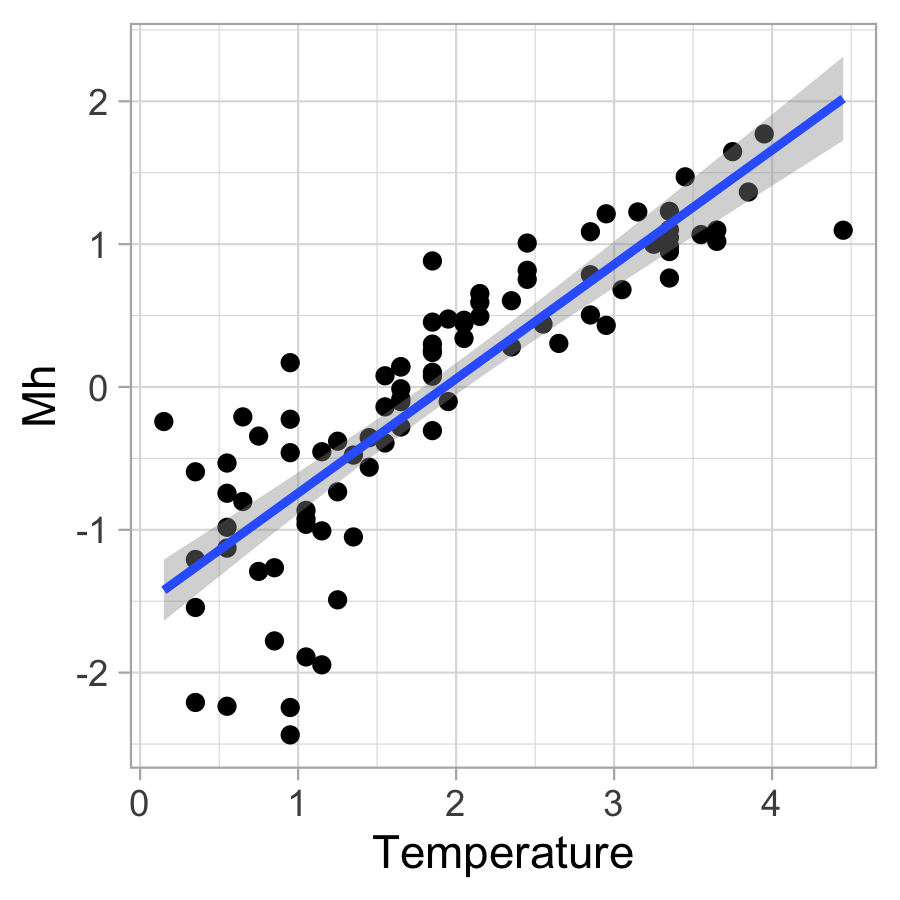}
    \caption{Missing actor estimated vector of means $M_h$ as a function of the temperature. $\corHTemp=0.85$.}
    \label{fig:barents_temp}
\end{figure}

In terms of interpretation, Figure \ref{fig:barents_temp} shows that the missing actor is highly correlated with the temperature. It also appears that the abundances of the species neighbor to the missing actor are much more correlated with the temperature (mean correlation = 0.78, sd = .06) than the abundances of the non-neighbor species (mean correlation = 0.46, sd = .27). This example shows the ability of the method to recover an underlying effect that would not be recorded in the data.

\subsection{Fatala River}
\begin{figure}[h]
    \centering
    \includegraphics[width=12cm]{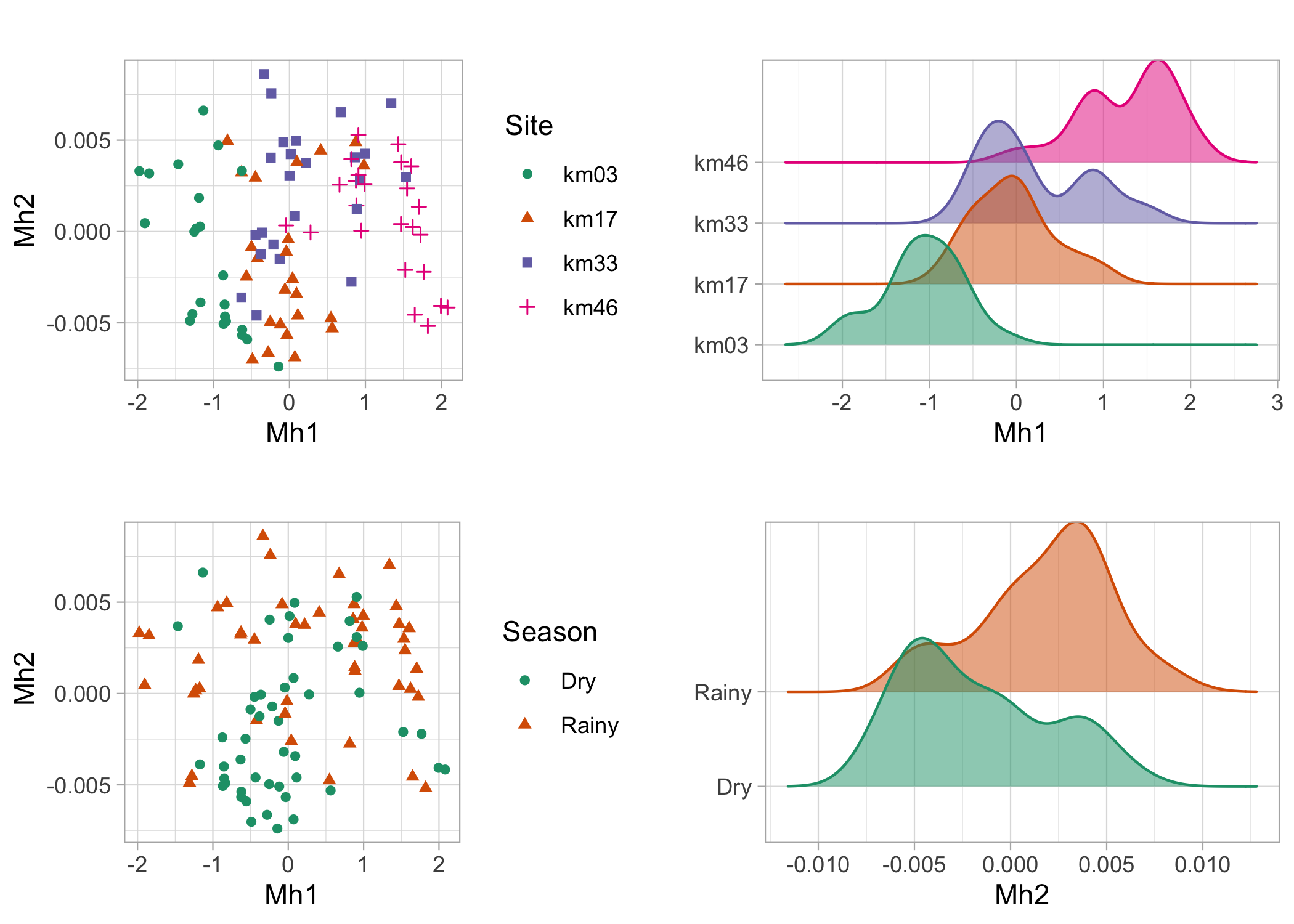}
    \caption{Estimated means $M_{h_1}$ and $M_{h_2}$ of the two inferred missing actors. Left column: scatterplots $M_{h_1}$ vs $M_{h_2}$ with site (top) and season (bottom) color code. Right: distribution of the estimated means across sites. Top right: distribution of $M_{h_1}$ in each location, bottom right: distribution of $M_{h_2}$ in each season.}
    \label{fig:Fatala}
\end{figure}

\cite{baran1995dynamique} collected the abundances of 33 fish species in 90 sites along the Fatala River in Guinea between June 1993 and February 1994. The data are available from the R package \texttt{ade4} on CRAN \citep{dray2007ade4}, along with the date and site of collection, from which we deduce the season (dry or rainy). Again the model was fitted without any covariates, but with two missing actors, as suggested by Figure \ref{fig:selec}. \\
The resampling initialization procedure yielded in 60 different cliques, for each of which a VEM algorithm was run: the mean running time was $11.33$ min (sd = $1.47$ mn). 14 VEM did not reach convergence (with tolerance $\varepsilon = 1e-3$) after 100 iterations. We filtered out the results obtained from the different initializations, when the algorithm obviously ended in a degenerate solution ($\Var(M_h) < \exp(-20)$). \\ 
%
Figure \ref{fig:Fatala} shows the scatterplot of the estimated conditional mean of the two missing actors $(M_{h_1}, M_{h_2})$ in each site, colored with either one of the available covariates (site and season). The missing actor $h_1$ is obviously linked to the site and separates most upstream locations (kilometer 3) from most downstream locations (kilometer 46). This actor has 11 highly probable neighbor species.
Again, this retrieved missing actor corresponds to an underlying effect (in this case: geography) that rules fish species abundances. \\
The second missing actor seems to be linked with the season but with a less clear separation. Also the variability of $M_{h_2}$ is much smaller than this of $M_{h_1}$. This effect is therefore questionable, which brings us back to model selection. As mentioned above, we used a procedure based on cross-validation, which may  be prone to select too complex model \citep{shao1993linear,friedman2001elements,arlot2010survey}. The definition of a grounded model selection criterion for structure inference in presence of missing actors remains open.

 
\paragraph{Acknowledgements.} 
This work was partly supported by the French ANR-18-CE02-0010 Ecological Networks (EcoNet) project and by the French ANR-11-LABX-0056-LMH LabEx Laboratoire de Mathématique Hadamard.

\begin{appendix}
 \section{Algebraic Tools} \label{app:tools}
 We here present some algebraic results about spanning tree structures which are used during the computations. Theorem \ref{thm:MTT}, Lemma \ref{lem:Meila} as well as Lemma \ref{lem:Kirshner} use the notion of Laplacian matrix  $\Qbf$ of a symmetric matrix $\Wbf=[w_{jk} ]_{1\leq j,k\leq p}$, which is defined as follows :
 
\[
 [\Qbf]_{jk}  =\begin{cases}
    -w_{jk}  & 1\leq j<k \leq p\\
    \sum_{u=1}^p w_{ju} & 1\leq j=k \leq p.
    \end{cases}
\]
 
We further denote $\Wbf^{uv}$ the matrix $\Wbf$ deprived from its $u$th row and $v$th column and we remind that the $(u, v)$-minor of $\Wbf$ is the determinant of this deprived matrix, that is $|\Wbf^{uv}|$.
The following Theorem \ref{thm:MTT} is the extension of Kirchhoff's Theorem to the case of weighted graphs \citep{matrixtree,MeilaJaak}.\\
\begin{theorem}[Matrix Tree Theorem] \label{thm:MTT}
    For any symmetric weight matrix W with all positive entries, the sum over all spanning trees of the product of the weights of their edges is equal to any minor of its Laplacian. That is, for any $1 \leq u, v \leq p$,
   \[
    W := \sum_{T\in\mathcal{T}} \prod_{(j, k)\in T} w_{jk} = |\Qbf^{uv}|.
    \]\\
\end{theorem}    

In the following, without loss of generality, we will choose $\Qbf^{11}$. As an extension of this result, \cite{MeilaJaak} provide a close form expression for the derivative of $W$ with respect to each entry of $\Wbf$. 

\begin{lemma} [\cite{MeilaJaak}] \label{lem:Meila}
    Define the entries of the symmetric matrix $\Mbf$ as
 \[    
 [\Mbf]_{jk} =\begin{cases}
    \left[(\Qbf^{11})^{-1}\right]_{jj} + \left[(\Qbf^{11})^{-1}\right]_{kk} -2\left[(\Qbf^{11})^{-1}\right]_{jk} & 1< j<k \leq p\\
    \left[(\Qbf^{11})^{-1}\right]_{jj} & k=1, 1< j \leq p  \\
    0 &  j=k .
    \end{cases}
\]
it then holds that $$\partial_{w_{jk}} W = [\Mbf]_{jk}  \times W.$$\\
\end{lemma}

\cite{kirshner} build on Lemma \ref{lem:Meila} to provide an efficient computation of all edges probabilities.
\begin{lemma} [\cite{kirshner}] \label{lem:Kirshner}
    Let $p_W$ be a distribution on the space of spanning trees, such that $p_W(T)=\prod_{kl\in T} w_{kl} / W$, where $W$ is defined as in Theorem \ref{thm:MTT}. Taking the symmetric matrix $\Mbf$ as defined in Lemma  \ref{lem:Meila}, the probability for an edge $kl$ to be in the tree $T^*$ writes:
 
$$\mathds{P}\{kl\in T^*\} = \sum_{T\in \mathcal{T}} p_W(T)= w_{kl}\: \Mbf_{kl}$$
\end{lemma}

 \newpage

\section{Computations} \label{app:comput}
\subsection[Update of tree parameter vector]{Update of $\betabf$.} \label{up:beta}
As in \citet{MRA20}, the update of $\betabf$ is such that:
$$\betabf^{t+1}  = \arg\max_\betabf \; \Esp_{g^t} \left[ \log p_\betabf(T) \right].
$$
By definition of $p_\betabf(T)$:
$$\Esp_{g^t} \left[ \log p_\betabf(T) \right] = \sum_{kl} P^t_{kl} \log \beta_{kl} - \log B\;,
\qquad
B=\sum_{T\in \mathcal{T}}\prod_{kl\in T} \beta_{kl}.$$
Computing the derivative with respect to the edge weight $\beta_{kl}$ gives:
\begin{align*}
\partial_{\beta_{kl}}\Esp_{g^t} \left[ \log p_\betabf(T) \right] &=\frac{P_{kl}^t}{\beta_{kl}} - \frac{\partial_{\beta_{kl}} B^t }{B^t} 
\end{align*}
According to Lemma \ref{lem:Meila}: $\partial_{\beta_{kl}} B^t  = [\boldsymbol{M}]_{kl} \times B$. Finally setting the derivative to 0 yields the update formula $
\beta^{t+1}_{kl} 
= \frac{P^t_{kl}}{ M(\betabf^t)_{kl}}$.

\subsection[Update of Gaussian tree precision matrix]{Update of $\Omega_T$} \label{up:omega}
The update of $\Omegabf_T$ respects
$$\Omegabf^{t+1}  = \arg\max_\Omegabf \; \Esp_{q^t} \left[ \log p_{\Omegabf}(\Ubf \mid T) \right].$$
This is a problem of parameter optimisation in the context of Gaussian Graphical Models (GGM).
In what follows, for any $q\times q$  matrix $A$, $A_{[kl]}$ will refer to the bloc $kl$ of $A$: $A_{[kl]}=(a_{ij})_{\{i,j\}\in\{k,l\}}$.   $[A_{[kl]}]^q$ will then denote the matrix obtained by filling up with zero entries to obtain full dimension $q\times q$, so that:
$$([A_{[kl]}]^q )_{ij}=\left\{ \begin{array}{rl}
a_{ij} & \text{if } \{i,j\}\in\{k,l\}\\
0 &  \text{if } \{i,j\}\in\{1,..., q\}_{\setminus kl}
\end{array}\right.$$
In its proposition 5.9, \citet{Lau96} states that in a  GGM with $p$ variables and associated with the decomposable graph $\mathcal{G}$, the maximum likelihood of the precision matrix exists if and only if $n > \max_{C\in \mathcal{C}} |C|$. It is then given as 
$$\widehat{\Omega}=n\left(\sum_{C\in \mathcal{C}} [SSD_{[C]}\,^{-1}]^p - \sum_{S\in \mathcal{S}} \nu(S)\,[SSD_{[S]}\,^{-1}]^p \right)$$
where $\mathcal{C}$ is the set of cliques and $\mathcal{S}$ the set of separators of $\mathcal{G}$, with associated multiplicities $\nu(S)$.\\

In our context, $\mathcal{G}$ is a spanning tree and so all cliques are edges and separators are nodes. The multiplicity of a given node $k$ as a separator in the graph is  $\nu(k) = d(k)-1$, where $d(k)$ is its degree. Therefore the estimator  $\widehat{\Omega}_T$  writes as the following 
\begin{align*}
\widehat{\Omega}_T &= n  \sum_{kl\in T}   [(SSD_{[kl]})^{-1}]^{p+r} - n\sum_k (d(k)-1)[(SSD_{kk})^{-1}]^{p+r}\\
&=n \sum_{kl\in T}  [(SSD_{[kl]})^{-1} - (SSD_{kk})^{-1} -  (SSD_{ll})^{-1} ]^{p+r} + n\sum_k[(SSD_{kk})^{-1}]^{p+r}
\end{align*}
As $SSD$ has diagonal $n$, the expression simplifies. Denoting $I_d$ the identity matrix of dimension $d$ we obtain:
$$\widehat{\Omega}_T =n\sum_{kl\in T} [(SSD_{[kl]})^{-1} -\frac{1}{n} I_2]^{p+r}+ I_{p+r}.$$

Detailing each bloc matrices as follows gives the update formulas in (\ref{omegaT}):
\[
n\times [(SSD_{[kl]})^{-1} - \frac{1}{n}I_2] = \frac{1}{1-(ssd_{kl}/n)^2}
\left(\begin{array}{cc}
		(ssd_{kl}/n)^2   & -ssd_{kl}/n\\
		-ssd_{kl}/n& (ssd_{kl}/n)^2 
		\end{array}\right)
\]

\subsection[for the toc]{Determinant of $\Omegabf_T$.}
The determinant of a precision matrix of a GGM with a decomposable graph is expressed as follows \citep{Lau96}:
$$ |\Omega| =\dfrac{\prod_{C\in \mathcal{C}} |\Sigma_C|^{-1}}{\prod_{S\in \mathcal{S}} |\Sigma_S|^{-\nu(S)}},$$
where $\Sigma = \Omega^{-1}$. As $\Omegabf_T$ is tree-structured, its determinant factorizes on the edges of $T$. It is expressed with the correlation matrix $\Rbf_T$ as follows, denoting $d(k)$ the degree of node $k$:
\begin{align*}
|{\Omegabf}_T| &=\frac{\prod_{kl \in T} |{\Rbf}_{Tkl}|^{-1}}{\prod_k |{\Rbf}_{Tkk}|^{1-d(k)}} 
 \end{align*}
Using that $\Rbf_T$ has diagonal 1, we obtain for step $t+1$ of the algorithm:
$$|\Omegabf^{t+1}_{T}| = \Big(\prod_{kl \in T} |\Rbf_{T[kl]}^{t+1}|\Big)^{-1}.$$

\subsection{Numerical issues.} \label{app:numIssues}

\subsubsection*{Exact computations} Our algorithm requires the computation of determinants (from the Matrix Tree Theorem) and inverses (in Kirshner's formula) of Laplacian of weight matrices. As we deal with highly variable weights, numerical issues arise: infinite determinants or matrix numerically non-invertible due to either the maximal machine precision (about $1.7\cdot 10^{308}$), or with machine zero (about $2.2 \cdot 10^{-16}$). To enhance the precision of such computations, we rely on multiple-precision arithmetic which allows the digit of precision of numbers to be  limited only by the available memory instead of 64 bits. We implemented matrix inversion and log-determinant computation using both, symbolic computation and multiple precision arithmetic, relying on the \texttt{gmp} R package available on CRAN, which uses \citep{lucas2020package}, the C library GMP (GNU Multiple Precision Arithmetic). 

\subsubsection*{Tempering parameter $\alpha$} \label{alpha}
Weights $\widetilde{\beta}$ are mechanically linked to the quantity of data available $n$. To avoid reaching maximal precision when computing the determinant, a tempering parameter $\alpha$ is applied to every quantity proportional to $n$, so that the actual update performed is $$\log \widetilde{\beta}_{kl} = \log \beta_{kl} - \alpha(\frac{n}{2}\log|\widehat{\Rbf}_{Tkl}| + \widehat{\omega}_{Tkl} [M^\intercal M]_{kl}).$$
We provide hereafter a heuristic to set the parameter $\alpha$.

\paragraph{An upper bound for $\alpha$:} 
The proposed algorithm requires the computation of the normalizing constant $\widetilde{B}$, which is the determinant of any minor of the Laplacian  of the $q\times q$ variational weights matrix $\betabft$. As these weights  mechanically increase with the quantity of available data $n$, this step is numerically very sensitive.  Hereafter we denote $|\Qbf^{uv}|$ this determinant and $\Delta$ the maximal machine precision. In order to ease the computations, we define the tempering parameter $\alpha$ as $$\log \widetilde{\beta}_{kl} = \log \beta_{kl} - \alpha(\frac{n}{2}\log|\widehat{\Rbf}_{Tkl}| + \widehat{\omega}_{Tkl} [M^\intercal M]_{kl})\;,\qquad \text{under constraint}\;\;\; |\Qbf^{uv}| \leq \Delta.$$

Let's first detail the expression for $\widetilde{\beta}_{kl}$. Following the definition of the $SSD$ matrix, and update formulas \eqref{omegaT} and \eqref{RT}, we obtain:
\begin{align*}
    \log \widetilde{\beta}_{kl} &=\log \beta_{kl} +\alpha \,n\left\{\frac{(ssd_{kl}/n)^2}{1-(ssd_{kl}/n)^2} -\frac12\log\big[1-(ssd_{kl}/n)^2\big]\right\}
\end{align*}
For large $n$, we thus have $$\widetilde{\beta}_{kl}\approx \exp \big[\alpha n \cdot C(ssd_{kl}/n)\big], \qquad \text{with }\; C(x)=x/(1-x) -\log(\sqrt{1-x}),\; x\in [0,1[.$$ 
We then define $C_{sup}$ such that $C_{sup} = C(ssd_{max})$, with $ ssd_{max}=\max\{ssd_{kl}, k\neq l\}$.
By definition, $\Qbf^{uv}$ is positive-definite, so its determinant is upper bounded by the product of its diagonal terms (Hadamard's inequality). Namely:
\begin{align*}
    |\Qbf^{uv}|&\leq \prod_{i=1}^{q-1} \Qbf^{uv}_{ii} \leq \prod_{i=1}^{q-1}\sum_{i=1}^{q-1} \exp (\alpha C_{sup} n)\\
    &\leq \left[(q-1)\exp(\alpha C_{sup} n)\right]^{q-1}
\end{align*}
Then applying the constraint yields:
\begin{align*}
    |\Qbf^{uv}| \leq \Delta \iff  \alpha \leq \frac{1}{C_{sup} n} \left[ \frac{1}{q-1}\log \Delta - \log(q-1)\right] 
\end{align*}

For $C_{sup}=0.8$, $n=200$ and $q=15$, we get $\alpha \leq 1.05\cdot 10^{-1}$.
 \newpage
\section{Model selection and cross-validation} \label{sec:modSel}

\subsection{Sampling spanning trees} \label{eq:sampTree}
Sampling non-uniform spanning trees (i.e. sampling $T$ from $p_\betabf$) is a research topic by itself, especially for large networks \citep[see][for a review]{DKP17}. For moderate size networks, a rejection algorithm \citep{Dev86} can be defined in the following way:
\begin{enumerate}
\item Sample $T$ from a distribution $q$, such that there exist a constant $M$, that ensures that, for all $T$, $M q(T) > p_\betabf(T)$;
\item Keep $T$ with probability $M^{-1} p_\betabf(T) / q(T)$ or try step 1 again.
\end{enumerate}
The efficiency of such an algorithm strongly relies on the choice of the proposal distribution. Here we adopt the following proposal:
\begin{enumerate}[($i$)]
\item Sample a connected graph $G$ with independent edges, each drawn with probability $Q_{jk} \propto P_{jk} = \Pr_\betabf\{ jk \in T\}$; 
\item Sample $T$ uniformly among the spanning trees of $G$.
\end{enumerate}
\paragraph{Evaluation of the proposal.}
To evaluate the proposal distribution for each sampled tree, we may observe that, the probability for a graph drawn from the proposal to contain a given tree $T$ is approximately
$$
{\Pr}_q\{G \ni T\} \approx \prod_{jk \in T} Q_{jk},
$$
the approximation being due to the connectivity constraint. This constraint can be almost surely satisfied by taking $Q_{jk}$'s large enough. So, denoting $|\Tcal(G)|$ the number of spanning trees in $G$, we have that
\begin{align*}
q(T) 
= \sum_{G \ni T} q(T \mid G) q(G)  = \sum_{G \ni T} \frac{q(G)}{|\Tcal(G)|} 
= {\Pr}_q\{G \ni T\} \; \Esp\left(|\Tcal(G)|^{-1} \mid G \ni T \right).
\end{align*}
The last expectation can be evaluated via Monte-Carlo, by sampling a series of graphs $G$ according to the proposal $q$ but forcing all edges from $T$ to be part of $G$. 
\paragraph{Upper bounding constant $M$.}
To evaluate the upper bounding constant $M$, we may observe that finding the tree $T^*$ such that
$$
m_\betabf 
:= \frac{{\Pr}_q\{G \ni T^*\}}{p_\betabf(T^*)}
= \min_{T \in \Tcal} \frac{{\Pr}_q\{G \ni T\}}{p_\betabf(T)} = \min_{T \in \Tcal} \prod_{jk \in T} \frac{Q_{jk}}{\beta_{jk}}
$$
is a minimum spanning tree problem. Then, obviously, for any tree $T$: ${\Pr}_q\{G \ni T\} \geq m_\betabf p_\betabf(T)$.
Now, because the maximum number of spanning trees within a graph is $p^{p-2}$, we have
$$
M q(T)
= M \sum_{G \ni T} \frac{q(G)}{|\Tcal(G)|} 
\geq \frac{M}{p^{p-2}} \sum_{G \ni T} q(G)
= \frac{M}{p^{p-2}} {\Pr}_q\{G \ni T\}
\geq M \frac{m_\betabf}{p^{p-2}}  p_\betabf(T)
$$
So we may set $M = p^{p-2} / m_\betabf$. Still, in practice, this bounds turns out to be far too large and needs to be tuned down to preserve computational efficiency.

 
\subsection{Cross-validation for model selection} \label{eq:cvAlgo}
\label{CV}

The cross-validation procedure to estimate the pairwise composite likelihood is given in Algorithm \ref{algo:model-selection}. In practice $V=10$ and $B = 100$.

\begin{algorithm}
\caption{Cross-validation for model selection with $r$ missing actors}
\label{algo:model-selection}
  \CommentSty{// 0. INITIALIZATION}\; 
  Divide the dataset $\Ybf$ into $V$ subset $\Ybf^1, \dots \Ybf^V$;
  \BlankLine
  \For{$v \in \{1,\cdots, V\}$}{
    \BlankLine
    \CommentSty{// 1.   Apply the VEM algorithm to the train dataset $\Ybf^{-v}$}\; 
    $\Gammabf_r^{-v} \leftarrow (\thetabf_r^{-v}, \sigmabf_r^{-v}, \betabf^{-v}_r, \Omegabf_r^{-v})$
    \CommentSty{// 2. MONTE CARLO APPROXIMATION OF COMPLETE LOG-LIKELIHOOD EXPECTATION}\;
    \For{$b \in \{1,\cdots, B\}$}{
    \CommentSty{// 2.1 Draw tree (see Section \ref{eq:sampTree})}\; 
     $ T_{r, b}^{-v}  \sim p_{\betabf^{-v}_r}$ 
    \BlankLine
     \CommentSty{// 2.2. Build  the precision matrix having non-nul entries determined by $ T_{r, b}^{-v} $ and values stored in $\Omegabf_r^{-v}$, and its diagonal terms according to \eqref{omegaT}}\;
     $\Omegabf_{T^b}\leftarrow f( T_{r, b}^{-v} , \Omegabf_r^{-v} )$
     \BlankLine
      \CommentSty{// 2.3. Compute the marginal variance matrix }\;
      $\Sigmabf_{T^bO} \leftarrow  \Omegabf_{T^bOO} - \Omegabf_{T^bOH} \Omegabf_{T^bHH}^{-1} \Omegabf_{T^bHO}$;
          \BlankLine 
     \CommentSty{// 2.4. Compute the bivariate Poisson log-normal density in test sites}\;
     \For{site $i \in v$}{
        \For{pairs of species $(j,k)$}{
        $p_{PLN}\left((Y^v_{ij}, Y^v_{ik}); \Gammabf_r^{-v}, T_{r, b}^{-v} \right)$ with means $\xbf_i^\intercal \thetabf_{r, j}^{-v}$ and $\xbf_i^\intercal \thetabf_{r, k}^{-v}$ and variance matrix $[\Sigmabf_{T^bO}]_{[jk, jk]}$
         }}
       \CommentSty{// 2.5.  Compute the average}\;
        $$
        PCL_{rvb}(\Ybf^v, \Gammabf_r^{-v}, T^b) = \frac1{m_v} \sum_{i = 1}^{m_v} \sum_{j < k} \log p_{PLN}\left((Y^v_{ij}, Y^v_{ik}); \Gammabf_r^{-v}, T_{r, b}^{-v} \right)
        $$
    }
    \BlankLine        
  }  
   \CommentSty{// 3. AVERAGE OVER SUBSETS}\; 
$$
PCL_r(\Ybf) = \frac1V \sum_v PCL_{rv}(\Ybf^v, \Gammabf_r^{-v}) .
$$
  \BlankLine
\end{algorithm}

\end{appendix}

\end{document}